\documentclass[aps,prl,showpacs,twocolumn,superscriptaddress]{revtex4-1}

\usepackage{dcolumn}
\usepackage{bm}
\usepackage[dvipdfmx]{graphicx}
\usepackage{color}
\usepackage{array}
\usepackage{mathrsfs}
\usepackage{ulem} 
\usepackage{amsmath}
\usepackage{amssymb}
\usepackage{url}

\usepackage{ulem} 

\begin{document}
\title{Resonating dimer-monomer liquid state in a magnetization plateau of a spin-$\frac{1}{2}$ kagome-strip Heisenberg chain}

\author{Katsuhiro Morita}
\email[e-mail:]{katsuhiro.morita@rs.tus.ac.jp}
\affiliation{Department of Applied Physics, Tokyo University of Science, Tokyo 125-8585, Japan}
\affiliation{Graduate School of Arts and Sciences, The University of Tokyo, Tokyo 153-8902, Japan}

\author{Shigetoshi Sota}
\affiliation{RIKEN Center for Computational Science (R-CCS), Kobe, Hyogo 650-0047, Japan}

\author{Takami Tohyama}
\affiliation{Department of Applied Physics, Tokyo University of Science, Tokyo 125-8585, Japan}

\date{\today}

\begin{abstract}
Highly frustrated spin systems such as the kagome lattice (KL) are a treasure trove of new quantum states with large entanglements.
We thus study the spin-$\frac{1}{2}$ Heisenberg model on a kagome-strip chain (KSC), which is one-dimensional KL, using the density-matrix renormalization group (DMRG) method.
Calculating central charge and entanglement spectrum for the KSC, we find a novel gapless spin liquid state with doubly degenerate entanglement spectra in a 1/5 magnetization plateau.
We also obtain a gapless low-lying continuum in the dynamic spin structure calculated by dynamical DMRG method.
We propose a resonating dimer-monomer liquid state that would meet these features.
\end{abstract}


\maketitle
Quantum entanglement is a very important concept for research such as quantum information and quantum magnetism~\cite{QE}.
Low-dimensional quantum spin systems are attracting attention owing to emerging their ground states with strong quantum entanglement. 
Using quantum entanglement analysis for quantum spin models has recently attracted attention because it is able to characterize various phases~\cite{EE1,EE2,EE3,EE4,EE5,ES1,ES2,ES3,ES4}, even in the states without the translational symmetry breaking and any long-range-dipole orders, such as the Haldane state in integer spin chains~\cite{Hald} and the Tomonaga-Luttinger liquid (TLL) state in half-integer spin chains.
According to the conformal field theory, the central charge $c$ charactering  excitation properties can be obtained by calculating the entanglement entropy (EE)~\cite{EE1,EE2,EE3,EE4,EE5}. 
In the TLL state, there is a gapless excitation with a non-zero integer value of $c$,  while in the Haldane state, there is a gapped excitation with $c=0$. 
Moreover, the Haldane state can be characterized by the degeneracy of entanglement spectrum (ES)~\cite{ES1,ES2,ES3,ES4}. For example, the spin-1 chain exhibits a doubly degenerate ES~\cite{ES1}. 

In two-dimensional frustrated quantum spin systems,
quantum spin liquid states such as resonating  valence  bond  (RVB) introduced by Anderson are expected to emerge~\cite{RVB1,RVB2}.
The RVB state is a resonant state of singlet dimers covering the whole lattice.
It is believed that one of possible models to exhibit the RVB states is the spin-$\frac{1}{2}$ antiferromagnetic Heisenberg model on the kagome lattice (KL)~\cite{KZ21}.
However, the ground state has been predicted to be $Z_2$ RVB spin liquid with topological order~\cite{KZ21,KZ22,KZ23,KZ24}, U(1) spin liquid~\cite{KU11,KU12,KU13,KU14}, and valence bond crystal (VBC)~\cite{VBC1,VBC2,VBC3,VBC4,VBC5}, although the exact ground state is unknown due to difficulty solving two-dimensional frustrated systems.
In the presence of magnetic field, magnetization plateaus have been predicted at $M/M_{\rm sat}$ = 0, 1/9, 1/3, 5/9, and 7/9, where $M_{\rm sat}$ is the saturation magnetization~\cite{KZ23,KMH1,KMH2,KMH3}.
These magnetization plateaus are realized by the effect of strong geometric frustration peculiar to the KL.

\begin{figure}[tb]
\includegraphics[width=70mm]{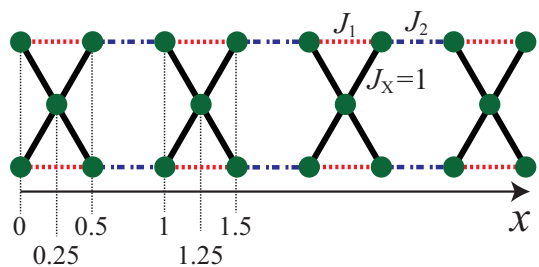}
\caption{Structure of a kagome-strip chain. The black solid, red dashed, and blue broken lines denote the exchange interactions $J_{\rm X}$, $J_1$, and $J_2$, respectively.
The numerics below the vertical dotted lines represent the distance of sites along the $x$ axis from the left edge.
We set $J_{\rm X}$=1.
\label{lattice}}
\end{figure}

Kagome-strip chains (KSCs), which are one-dimensional KL, have been recently studied, because the presence of exotic quantum states would be expected as in the KL~\cite{ksc1,ksc2,ksc3,ksc4,ksc5,ksc6,ksc7,ksc8}.
In KSC shown in Fig.~\ref{lattice}, it has been found that the magnetization plateaus emerge at $M/M_{\rm sat}$ = 0, 1/5, 3/10, 1/3, 2/5, 7/15, 3/5, and 4/5~\cite{ksc6}.
We can, therefore, expect the presence of novel quantum phases in these plateaus generated by strong geometrical frustration.

In this Letter, we investigate the 1/5 plateaus of the KSC using the density-matrix renormalization group (DMRG) method. We determine a magnetic phase diagram of the 1/5 plateau and we find two new plateau phases that have not been identified in the previous study~\cite{ksc6}. 
Our main result is that, even though one of the two new phases exhibits a gapless spin liquid behavior with $c=1$, the ES of the phase is doubly degenerate.
This means that this plateau phase has both properties of half-integer spin chain and spin-1 chain.
Furthermore, we calculate the dynamical spin structure factor (DSSF) in this phase by mean of the dynamical DMRG (DDMRG)~\cite{DDMRG}. 
The DSSF of the $S_z$ (magnetic field) direction exhibits gapless and dispersionless low-energy excitations.
In order to describe these novel properties in the new phase, we propose resonating dimer-monomer liquid (RDML) state that is a mixed state of singlet dimers and up-spin monomers.

\begin{figure}[tb]
\includegraphics[width=86mm]{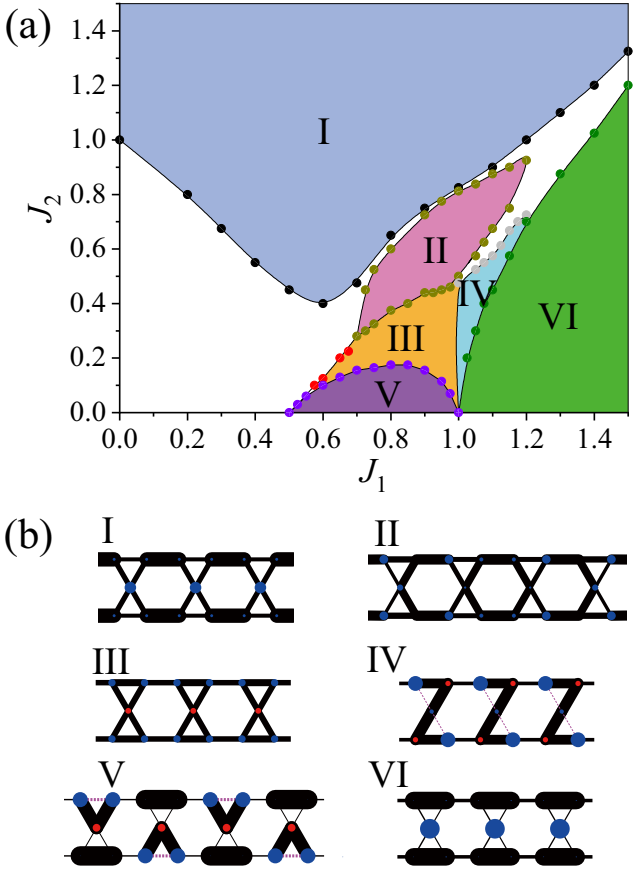}
\caption{(a) Phase diagram of the KSC at $M/M_{\rm sat}$=1/5.
Roman numbers denote different 1/5 plateau phases.
(b) The nearest-neighbor spin-spin correlation $\langle \mathbf{S}_i\cdot\mathbf{S}_j\rangle - \langle S^z_i\rangle\langle S^z_j\rangle$ and the local magnetization $\langle S^z_i\rangle$ around the center of the chain for the six phases. 
Black solid (purple dashed) lines connecting two nearest-neighbor sites denote negative (positive) values of the spin-spin correlation and their thickness represents the magnitude of correlation. 
Blue (red) circles on each site denote positive (negative) value of $\langle S^z_i\rangle$ and their diameter represents its magnitude.
\label{phase}}
\end{figure}

The Hamiltonian for the spin-$\frac{1}{2}$ KSC in magnetic field is defined as
\begin{eqnarray} 
H &=& \sum_{\langle i,j \rangle }J_{i,j} \mathbf{S}_i \cdot \mathbf{S}_j - h\sum_i S^{z}_i,
\end{eqnarray}
where $\mathbf{S}_i$ is the spin-$\frac{1}{2}$ operator, $\langle i,j \rangle$ runs over the nearest-neighbor spin pairs, $J_{i,j}$ corresponds to one of $J_{\rm X}$, $J_1$, and $J_2$ in Fig.~\ref{lattice}, and $h$ is the magnitude of magnetic field.  In the following we set $J_{\rm X}=1$ as energy unit.
We perform DMRG calculations at zero temperature up to system size $N=1000(=5\times200)$ for various values of $J_1$ and $J_2$. 
The number of states $m$ kept in the DMRG calculation are  $400-2500$ and truncation errors are less than $5 \times 10^{-7}$.
For the calculation of the dynamical spin structure factors, 
we use DDMRG (for details, see Ref.~\cite{DDMRG}) for $N=60$ and $N=80$ under the periodic boundary condition (PBC).
In our DDMRG calculations, $m$ is set to be 1000.

We first determine the phase diagram of the KSC at $M/M_{\rm sat}=1/5$.
We obtain six 1/5 plateau phases using $N=120-1000$ clusters under the open boundary condition (OBC) as shown in Fig.~\ref{phase}(a). 
The regions of the phases I-VI are denoted by different colors. The white region has no magnetization plateau.
The phases II and III are found in the present study, while other phases have already been found in the previous study~\cite{ksc6}.

\begin{figure}[tb]
\includegraphics[width=86mm]{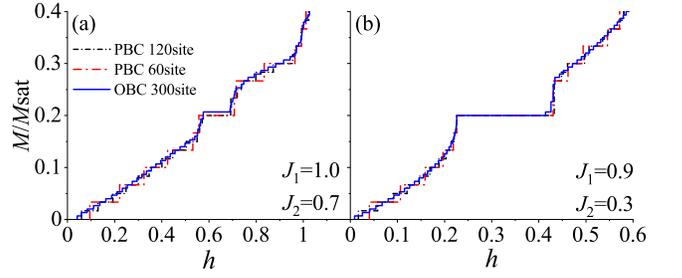}
\caption{Magnetization curve of the KSC under the PBC for $N=60$ and $N=120$ or the OBC for $N=300$ at zero temperature.
(a) $J_1=1.0$ and $J_2=0.7$ (phase II). (b) $J_1=0.9$ and $J_2=0.3$ (phase III). 
\label{M-H}}
\end{figure}

Figure~\ref{phase}(b)  shows nearest-neighbor spin-spin correlation $\langle \mathbf{S}_i\cdot\mathbf{S}_j\rangle - \langle S^z_i\rangle \langle S^z_j\rangle$ and local magnetization $\langle S^z_i\rangle$ for the six phases. The lines connecting two nearest-neighbor sites denote the sign and magnitude of spin-spin correlation by color and thickness, respectively. The circle on each site represents $\langle S^z_i\rangle$. 
The stability of the phases I, IV, V, and VI can be explained by energy gain due to local strong spin correlation as evidenced from thick lines in Fig.~\ref{phase}(b).
On the other hand, the phases II and III do not have such a distinguished thick line. 
In the phase III, the spin-spin correlation is almost uniform and there is no symmetry breaking.
In the phase II, periodic magnetic structure is not seen (see Supplemental Material~\cite{Suppl}).

Figure \ref{M-H} shows magnetization curves at $J_1=1.0$ and $J_2=0.9$ (phase II), and $J_1=0.9$ and $J_2=0.3$ (phase III).
The 1/5 plateaus are clearly visible in both conditions.
These magnetization curves show little variation with size $N$ and boundary conditions.
However, in Fig.~\ref{M-H} (a) (phase II), a plateau deviating from 1/5 just one step, that is, $M/M_{\rm sat}=1/5+2/N$, appears under the OBC.
We confirmed that this deviation appears in other sets of $J_1$ and $J_2$ on the phase II, 
implying the presence of edge excitations that appear in the Haldane chain.
The phases I, IV, V, and VI have been examined in the previous study. 
Therefore, we investigate the phases II and III in the following.

\begin{figure}[tb]
\includegraphics[width=86mm]{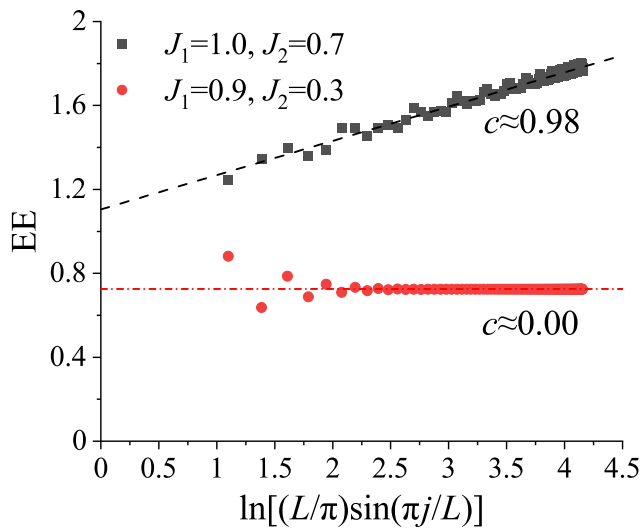}
\caption{Entanglement entropy of the KSC under the OBC for $L=200$ $(N=5\times200)$ at $J_1=1.0$, $J_2=0.7$ (Phase II) and $J_1=0.9$, $J_2=0.3$ (Phase III).
The value of the central charge $c$ is obtained by fitting of the EE data using straight lines. 
\label{EE}}
\end{figure}

Figure~\ref{EE} shows EE for $L=200$ $(N=5\times L)$ under the OBC as a function of $\ln[L/\pi \sin(\pi j/L)]$ with a variable $j$ denoting the position of the 5-site unit.
This plot comes from the following relation that holds between the central charge $c$ and the position-dependent EE, $EE(j)$,

\begin{eqnarray} 
EE(j) &=& \frac{c}{b_c}\ln\left[\frac{L}{\pi} \sin \left( \frac{\pi j}{L} \right) \right] + a_c,
\end{eqnarray}
where $a_c$ is a nonuniversal constant, and $b_c = 6 $ (3) for the OBC (PBC)~\cite{EE2,EE3,EE4,EE5}.
The value of $c$ becomes finite when the spin-spin correlation exhibits power-law decays, and gives $c=1$ in the TLL.
On the other hand, when spin-spin correlation decays exponentially, for example, in the Haldane phase, $c=0$.
The value of $c$ in Fig.~\ref{EE} is obtained by fitting the EE data using straight lines.
Since the $c$ for the phase II is nearly unity as in the TLL, the phase II is expected to be the gapless spin liquid. 
Since plateau phases should have an energy gap, there should be a gapless excitation only in the subspace where total $S_z$ does not change.
This feature has also been observed in the 1/3 magnetization plateau in frustrated three-leg spin tubes~\cite{TriT1,TriT2,TriT3}.
We also confirm that the spin-spin correlations of the $S^z$ and $S^x$ components exhibit power decay corresponding to the gapless excitation  and exponential decay corresponding to the energy gap, respectively (see Supplemental Material~\cite{Suppl}).

In the phase III, $c$ is nearly zero. 
This indicates that there is an energy gap in this phase. 
Moreover, we confirm that the ground state has no degeneracy and there is no symmetry breaking as shown in Fig.~\ref{phase}(b).

\begin{figure}[tb]
\includegraphics[width=86mm]{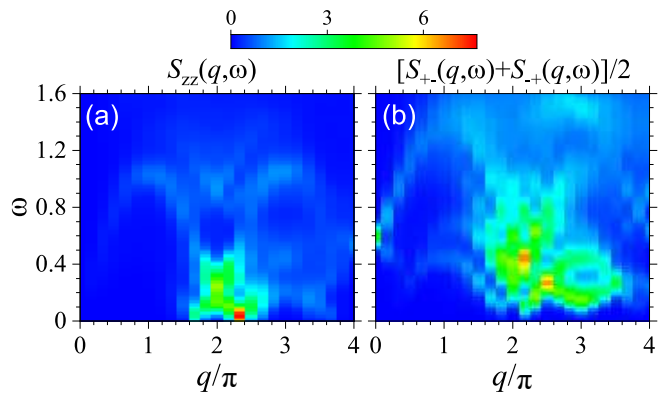}
\caption{Dynamical spin structure factor (a) $S_{zz}(q,\omega)$ and (b) $[S_{+-}(q,\omega)$+$S_{-+}(q,\omega)]/2$ obtained by the DDMRG for the KSC under the PBC for $L=12$ ($N=5\times12$) at $J_1=1.0$ and $J_2=0.7$ (phase II).
\label{DDMRG}}
\end{figure} 

In order to investigate the phase II in more detail, we calculate the DSSF, $S_{\alpha \beta}(q,\omega)$ defined by
\begin{equation} 
S_{\alpha \beta}(q,\omega)=-\frac{1}{\pi}\mathrm{Im} \left<0\right| \tilde{S}_{-q}^\beta \frac{1}{\omega-H+E_0+i\eta} \tilde{S}_q^\alpha \left|0\right>,
\end{equation}
where $q$ is the momentum for the lattice geometry shown in Fig.~\ref{lattice}, $\left|0\right>$ is the ground state with energy $E_0$, and
 $\eta$ is a broadening factor.
$\tilde{S}_q^{\alpha(\beta)}=S_q^{\alpha(\beta)}-\left<0\right|S_q^{\alpha(\beta)}\left|0\right>$, where
 $S_q^{\alpha(\beta)}=N^{-1/2}\sum_i e^{iqx_i} S_i^{\alpha(\beta)}$ with $x_i$ being the position of spin $i$ and $\alpha(\beta) = +,-,z$.

Figure~\ref{DDMRG} shows $S_{zz}(q,\omega)$ corresponding to the $S^z$ component and $[S_{+-}(q,\omega)$+$S_{-+}(q,\omega)]/2$ corresponding to the  $S^x$ and $S^y$ components for $L=12$ ($N=5\times12$) with $\eta=0.05$ under the PBC.
Since the position $x_i$ is defined in Fig.~\ref{lattice}, the range of $0\leq q/\pi \leq 4$ corresponds to a half of the extended Brillouin zone.
As shown in Fig.~\ref{DDMRG}(a), a gapless excitation in the $S^z$ component emerges around $q/\pi=2$. This excitation is consistent with the result expected from the fact that $c=1$.
We confirm that the value of $q$ showing the lowest-energy excitation is size dependent.
As the size is increased, the $q$ value tends to move toward $2\pi$ (see Supplemental Material \cite{Suppl}).
We thus expect that in the thermodynamic limit, a gapless excitation emerges at $q/\pi=2$.
In addition, low energy excitations at $\omega\lesssim0.2$ show less dispersive feature. 
This feature will be discussed later.

As shown in Fig.~\ref{DDMRG}(b), $[S_{+-}(q,\omega)$+$S_{-+}(q,\omega)]/2$ corresponding to the excitations of the $S^+$ and $S^-$ components, has a gap.
Here, the external magnetic field $h$ is set to be at the center of the 1/5 magnetization plateau.
The minimum excitation gap is found to exist around $q/\pi=2$ and $q/\pi=3$.
Continuous excitations exist up to the high energy region more than $\omega=1.5$ at $q/\pi=2$.

\begin{figure}[tb]
\includegraphics[width=86mm]{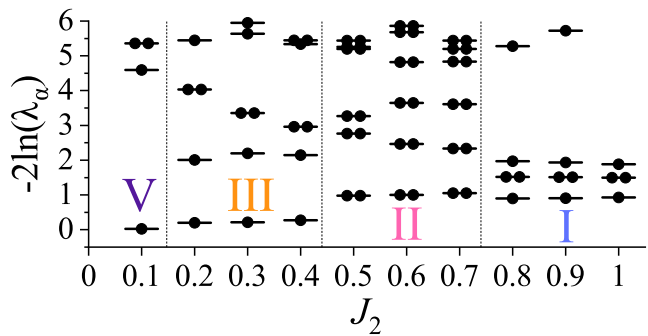}
\caption{Entanglement spectrum of the KSC under the OBC for $L=80$ $(N=5\times80)$ with respect to $J_2$ at $J_1=0.9$.
Roman numbers correspond to the phases in Fig.~\ref{phase}. The vertical dotted lines represent phase boundaries determined by the ES.
\label{ES}}
\end{figure}

Finally, we calculate the ES to investigate of each phase in more detail.
The ground state can be Schmidt decomposed as follows;
\begin{equation} 
\left|0\right>=\sum_{\alpha}\lambda_\alpha\left|\Phi^L_\alpha\right>\left|\Phi^R_\alpha\right>,
\end{equation}
where $\left|\Phi^L_\alpha\right>$ and $\left|\Phi^R_\alpha\right>$ are orthonormal basis
vectors of the left and right part of the chain, respectively~\cite{ES1}.
The $\lambda_\alpha^2$ are the eigenvalues of the reduced density matrix.
The ES is defined as $-2\ln(\lambda_\alpha)$.
Figure~\ref{ES} shows the results of the ES as a function of $J_2$ at $J_1=0.9$.
The phases I, III and V are trivial phases due to the mixture of singly and doubly degenerate states of the ES~\cite{ES1}.
Whereas, in the phase II, all the ES are doubly degenerate.
This is identical to the feature of the spin-1 Haldane chain. Therefore we conclude that the phase II is a non-trivial topological phase~\cite{ES1}. 
We also confirm that the double degeneracy is independent of the $L$, $J_1$, and $J_2$. The degeneracy is obtained even for $L=2$ (the shortest chain).

\begin{figure}[tb]
\includegraphics[width=86mm]{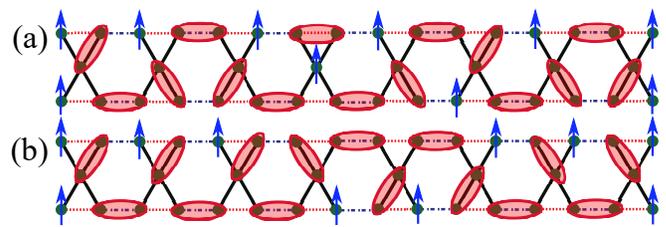}
\caption{Two snapshots of dimer-monomer states for $L=8$ under the OBC. 
All units except both ends in (a) and (b) have a monomer denoted by upward arrows. Dimers are schematically drawn by elliptic cone containing two sites.
\label{DML}}
\end{figure}

The phases II and III are newly identified magnetization plateau phases in the present study.
The phase III is concluded to be a trivial phase based on i) $c=0$, ii) no degeneracy in the ground state, and iii) trivial distribution of the ES.
On the other hand, the phase II is a novel phase that has the characteristics of both the gapless spin liquid and the spin-1 Haldane state, because of $c=1$ and the double degeneracy of the ES.
Why does the phase II have these characteristics?
We anticipate that a mixed state of singlet dimers and up-spin monomers, which reveals a liquid behavior like the RVB, is the ground state of the phase II.
We refer to this state as RDML.
The states in Fig.~\ref{DML} correspond to snapshots of the RDML state. 
They have one monomer in all units except at both ends, and there is no nearest-neighboring monomer.
The ES in these states shows double degeneracy. 
The RDML state whose typical components are shown in Fig.~\ref{DML} can generate infinitesimal energy excitations by swapping any dimers with monomers. This corresponds to local excitations, which form a dispersionless structure in the $S_{zz}(q,\omega)$.
The DMRG result shown in Fig.~\ref{DDMRG} confirms this property.
As the size is increased, the dispersionless excitations are more pronounced around $q/\pi=2$ (see Supplemental Material \cite{Suppl}).
Assuming the RDML state, we can explain the gapless exsitetion, double degeneracy of the ES, and dispersionless low-energy excitations in the $S_{zz}(q,\omega)$.
Therefore, we believe that the phase II is the RDML state.

In summary, we obtained the phase diagram of the 1/5 plateaus of the KSC using the DMRG method. 
We found two new plateau phases (II and III).
The most surprising result is that the phase II exhibits the gapless spin liquid with $c=1$ as in the half-integer spin chain and the double degeneracy of the ES as in the spin-1 chain.
We also calculated the DSSF in the phase II using the DDMRG. 
We found that the DSSF of the $S_{zz}(q,\omega)$ exhibits the gapless excitation corresponding to $c=1$ and dispersionless low-energy excitations corresponding to the local excitations.
Finally, we proposed the RDML state that can explain the gapless exsitetion, double degeneracy of the ES, and dispersionless low-energy excitations in the $S_{zz}(q,\omega)$ in the phase II.

The KSC compounds with five exchange interactions have already been reported~\cite{kscexp}. 
Therefore, there is a possibility that compounds with our model will be synthesized in the future. Since $M/M_{\rm sat}=1/5$ is a relatively low magnetization, the 1/5 plateau may be observed experimentally. 
Accordingly the experimental study of the KSC will be vigorously pursued in the future.

\begin{acknowledgments}
We acknowledge discussions with Gonzalo Alvarez from the Center For Nanophase Materials Sciences at ORNL.
This work was supported in part by MEXT as a social and scientific priority issue ``Creation of new functional devices and high-performance materials to support next-generation industries'' (CDMSI) to be tackled by using post-K computer and by MEXT HPCI Strategic Programs for Innovative Research (SPIRE) (hp190198). The numerical calculation was partly carried out at the K Computer and HOKUSAI, RIKEN Advanced Institute for Computational Science, the facilities of the Supercomputer Center, Institute for Solid State Physics, University of Tokyo, and the Information Technology Center, The University of Tokyo. This work was also supported by the Japan Society for the Promotion of Science, KAKENHI (Grants No. 19H01829 and No. JP19H05825)
\end{acknowledgments}

\newpage
\begin{widetext}
\begin{center}
\textbf{\large Supplementary Material for ``Resonating dimer-monomer liquid state in a magnetization plateau of a spin-$\frac{1}{2}$ kagome-strip Heisenberg chain''}\\
\ 
\\
Katsuhiro Morita,$^{1,2}$  Shigetoshi Sota,$^3$ and Takami Tohyama$^1$ \\
\vspace{1mm}
${ }^1$\textit{\small Department of Applied Physics, Tokyo University of Science, Tokyo 125-8585, Japan} \\
${ }^2$\textit{\small Graduate School of Arts and Sciences, The University of Tokyo, Tokyo 153-8902, Japan} \\
${ }^3$\textit{\small RIKEN Center for Computational Science (R-CCS), Kobe, Hyogo 650-0047, Japan} \\
\ 
\\
\end{center}
\end{widetext}
\setcounter{figure}{0} 
\renewcommand{\thefigure}{S\arabic{figure}}

\begin{figure}[b]
\includegraphics[width=86mm]{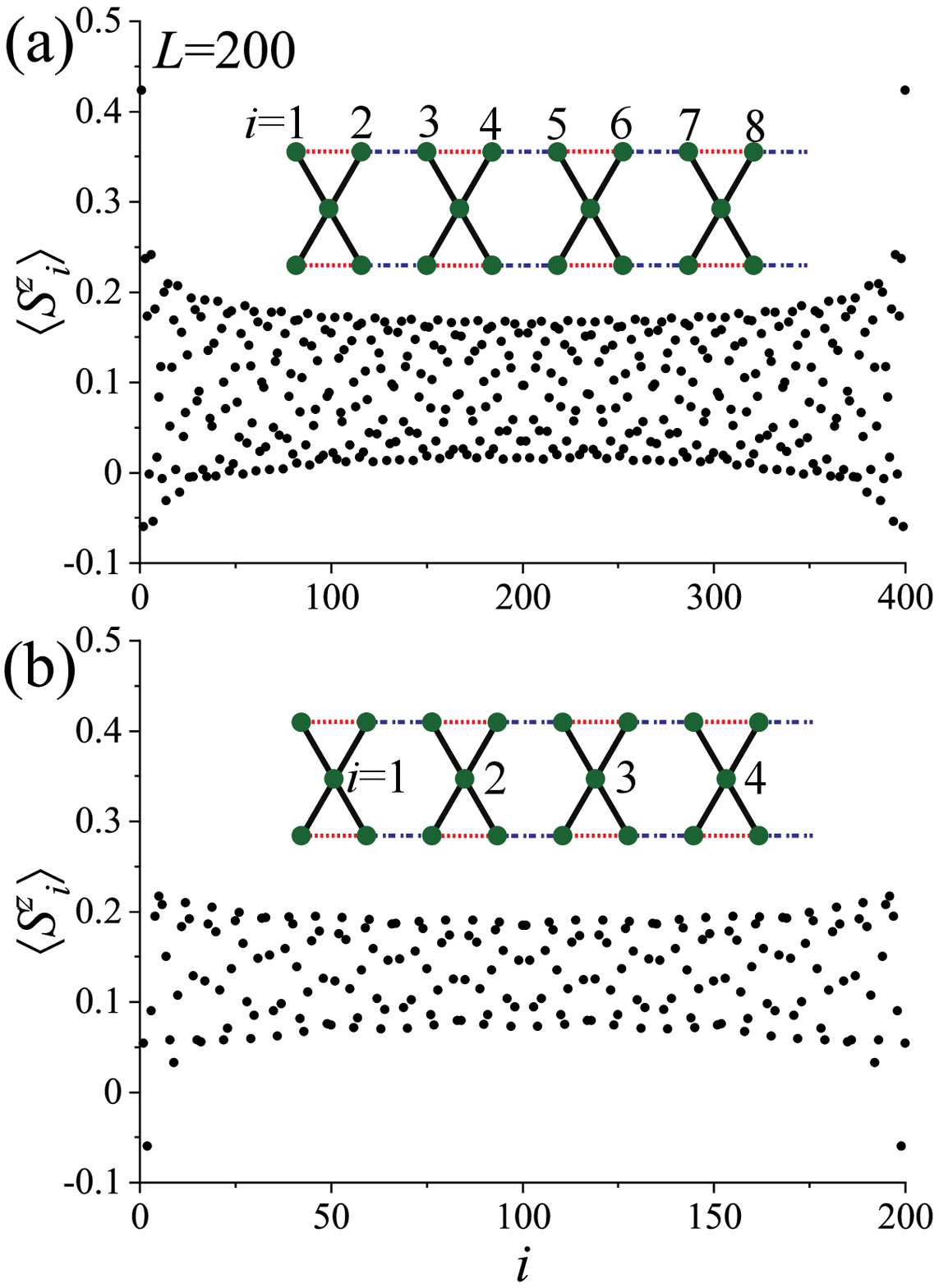}
\caption{Local moment $\left<S^z_i\right>$ at sites along the upper edge (a) and at central sites (b) of the KSC under the OBC for $L=200$ at $J_1=1.0$ and $J_2=0.7$. $i$ represents the position of each site labeled in the inset of each panel.
\label{Sz}}
\end{figure}

\begin{figure*}[tb]
\includegraphics[width=160mm]{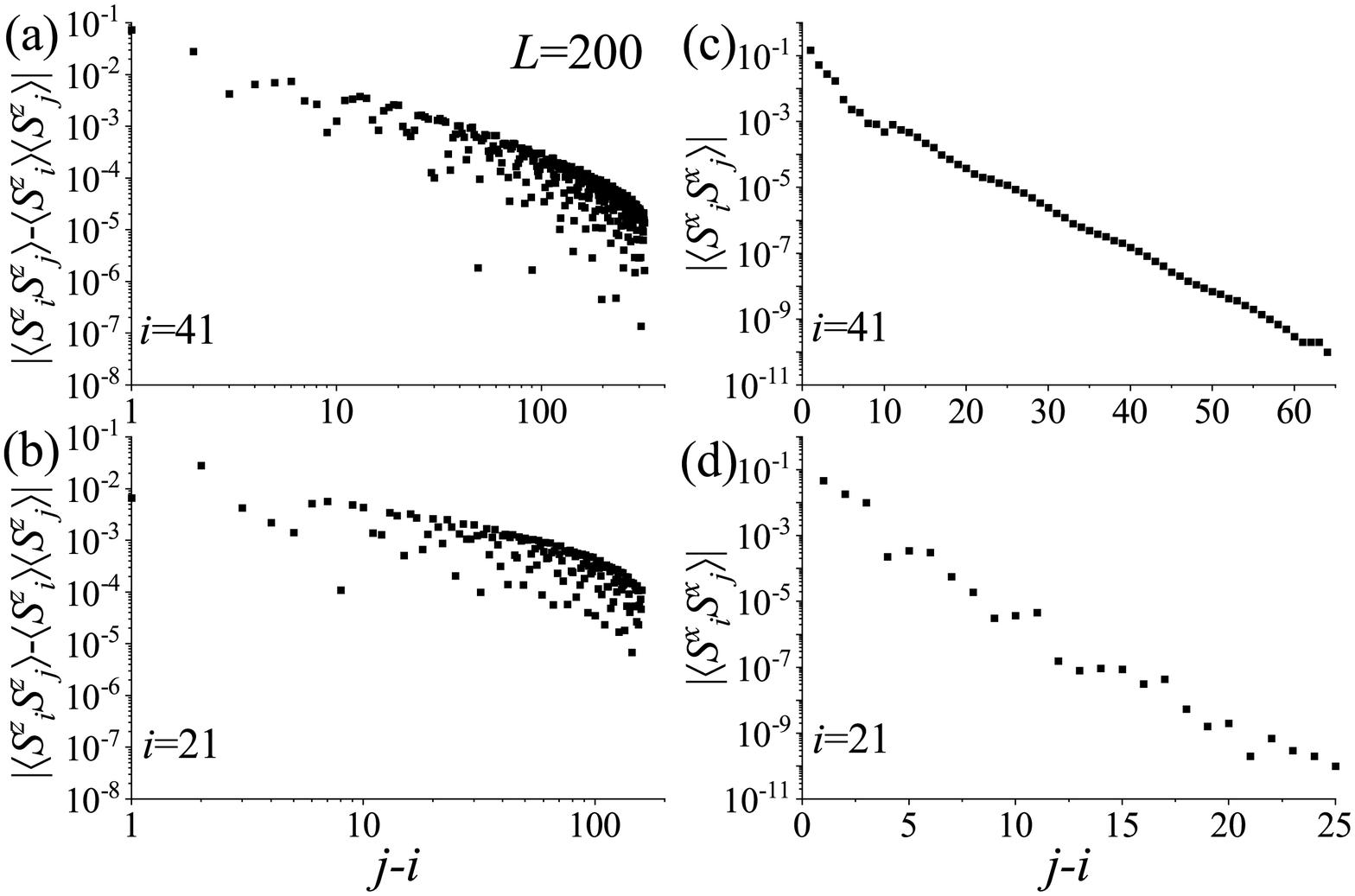}
\caption{ Spin-spin correlations of the KSC under the OBC for $L=200$ at $J_1=1.0$ and $J_2=0.7$.
(a)  $S_z$ component along sites on the upper edge.
(b)  $S_z$ component along the central sites.
(c)  $S_x$ component along sites on the upper edge.
(d)  $S_x$ component along the central sites.
The definition of $i$ and $j$ is the same as in Fig.~\ref{Sz}.
\label{SS}}
\end{figure*}

\begin{figure*}[tb]
\includegraphics[width=160mm]{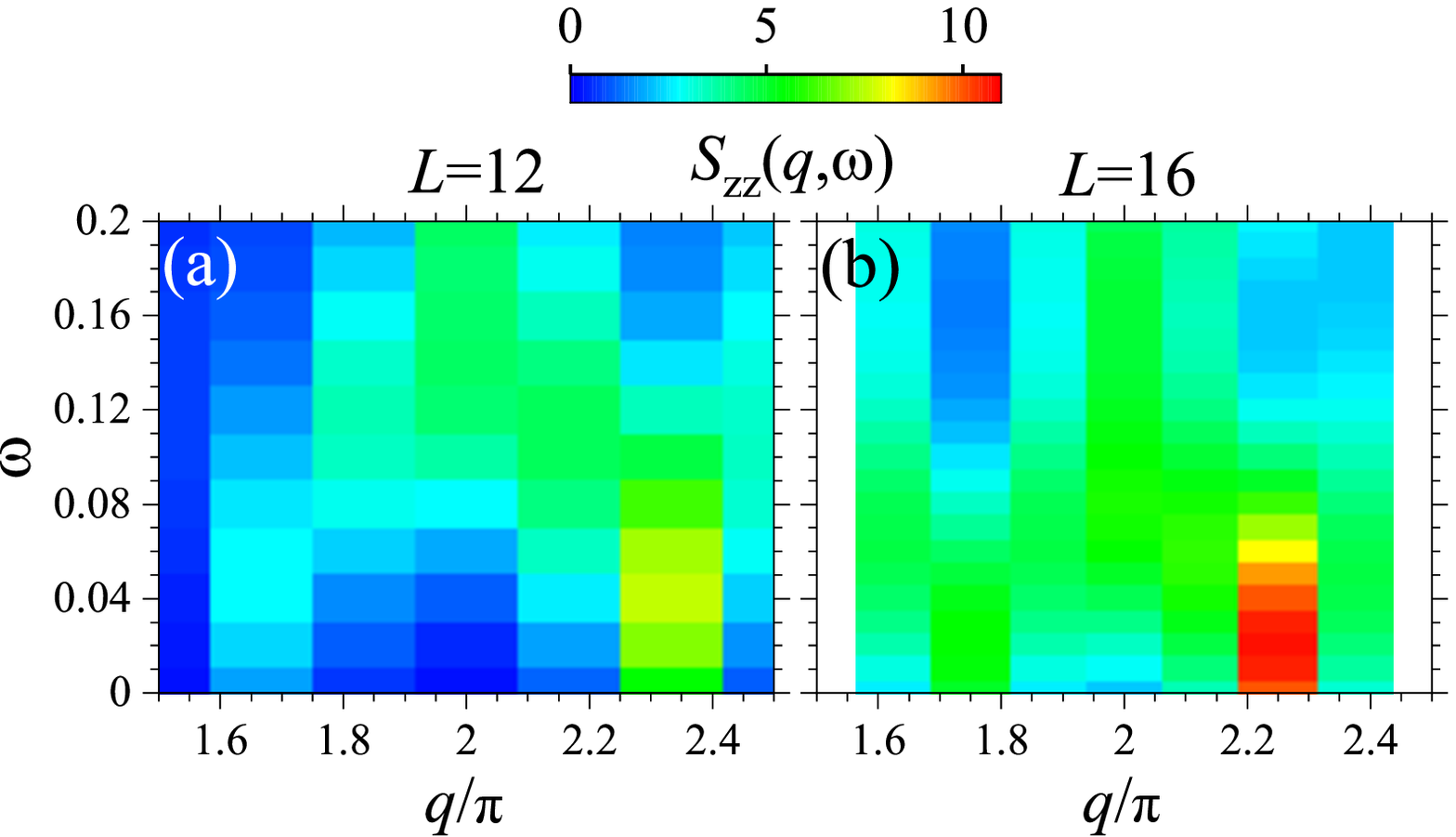}
\caption{ Dynamical spin structure factor $S_{zz}(q,\omega)$ obtained by the DDMRG for the KSC under the PBC  at $J_1=1.0$ and $J_2=0.7$ for (a) $L=12$ and (b) $L=16$.
\label{DDMRG2}}
\end{figure*}

Here, we show the results of the calculations of the local magnetization (LM) (Fig.~\ref{Sz}), spin-spin correlations (SSCs) (Fig.~\ref{SS}), and dynamical spin structure factor (DSSF) (Fig.~\ref{DDMRG2}) in the phase II at $J_1=1.0$ and $J_2=0.7$ (see Fig.~1 in the main text).
In our density-matrix renormalization group (DMRG) calculation for the LM and SSCs, we use the kagome-strip chain  (KSC) under the open boundary condition (OBC) for $L=200$ (total number of sites is $N=5\times L$; see Fig.~1 in the main text).
In order to obtain sufficiently accurate results, 
we set the number of states $m$ to be 2500 and resulting truncation error is less than $10^{-9}$.

Figures~\ref{Sz}(a) and \ref{Sz}(b) show the LM $\left<S_i^z\right>$ at sites along the upper edge and at central sites in the KSC, respectively.
Both ends of upper edge ($i=1$ and $i=400$) have almost full moment $\langle S_{i=1(400)}^z \rangle =0.42$, corresponding to the edge excitation.
We note that the LM at sites along the lower edge is the same as those along the upper edge plotted in Fig.~\ref{Sz}(a).
This means that there is no spontaneous breaking of the mirror symmetry with respect to the central axis along the chain.
Plateau phases usually have some long-range orders such as an up-up-down structure in the zigzag spin chain, while the phase II is completely different.
From the results in Figs.~\ref{Sz}(a) and~\ref{Sz}(b), we can see that the phase II is not a simple ordered state but a spin-density-wave-like state.

To investigate the phase II that has both gapless excitation and magnetization plateau in detail, we calculate the SSCs of the $S^z$ and $S^x$ components, which are defined by $\langle S^z_i S^z_j\rangle - \langle S^z_i\rangle \langle S^z_j\rangle$  and  $\langle S^x_i S^x_j\rangle$, respectively.
The SSCs of the $S^z$ and $S^x$ components exhibit power decay corresponding to the gapless excitation with $c=1$ as shown in Figs.~\ref{SS}(a) and~\ref{SS}(b) and exponential decay corresponding to the magnetization plateau as shown in Fig.~\ref{SS}(c) and~\ref{SS}(d), respectively.
These results are consistent with the fact that the phase II has $c=1$ and the magnetization plateau.

Finally, to confirm the finite size effect for the DSSF $S_{zz}(q,\omega)$, we calculate $S_{zz}(q,\omega)$ of the KSC under the periodic boundary condition (PBC) for $L=12$ and $L=16$ using dynamical DMRG (DDMRG).
Figure~\ref{DDMRG2} shows these results around $q/\pi=2$ in the low-energy region.
The value of $q$ showing lowest-energy excitation for $L=16$ ($2.25\pi$) is closer to $2\pi$ than that for $L=12$ ($2.33\pi$).
Furthermore, we find that low-energy excitations for $L=16$ are lower in energy and flatter than those for $L=12$.
We thus expect that in the thermodynamic limit, gapless excitation occurs at $q/\pi=2$ accompanied by flat-band excitations around $q/\pi=2$.

\end{document}